\documentstyle[aps,pra,twocolumn]{revtex}

\input{epsf}

\begin{document}
\draft
\title{Maximum likelihood estimation of photon number distribution\\
from homodyne statistics}
\author{Konrad Banaszek}
\address{Instytut Fizyki Teoretycznej,  Uniwersytet Warszawski,
Ho\.{z}a 69, PL--00--681 Warszawa, Poland}
\date{December 18, 1997}
\maketitle

\begin{abstract}
We present a method for reconstructing the photon number
distribution from the homodyne statistics based on maximization of the
likelihood function derived from the exact statistical description of a
homodyne experiment. This method incorporates in a natural way the physical
constraints on the reconstructed quantities, and the compensation for the
nonunit detection efficiency.
\end{abstract}


An interesting application of the homodyne detection is the measurement of
phase--insensitive properties of optical fields, dependent on the
photon number distribution \cite{MunrBoggPRA95}.  The homodyne
technique goes beyond the current limitations of direct photodetection.
First, ultrafast sampling time can be achieved by using the local
oscillator field in the form of a short pulse. Secondly, information on
the photon distribution is carried by two relatively intense fields,
which can be detected with substantially higher efficiency than the
signal field itself. This feature has enabled an experimental
demonstration of even--odd oscillations in the photon number
distribution of a squeezed vacuum state \cite{SchiBreiPRL96}.

In the homodyne scheme, all phase--independent properties of the measured
field are contained in the phase--averaged statistics of difference
counts \cite{BanaWodkPRA97}. 
The probability distribution of difference counts is a linear
combination of diagonal elements of the density matrix in the Fock
basis. This relation can be analytically inverted
\cite{DAriMaccPRA94},  which yields an expression for the photon number
distribution as integrals of the homodyne statistics with the so--called
pattern functions \cite{LeonMunrOpC96}. In a real experiment, however,
the homodyne statistics is known only with some statistical
uncertainty, as it is determined from a finite number of experimental
runs.  Application of pattern functions to noisy experimental data can
generate unphysical results, such as negativities in the photon number
distribution. These artifacts become particularly strong, when
compensation for the detector imperfection is used in the numerical
processing of the experimental data \cite{BreiSchiJMO97}.

In this communication we apply the statistical methodology of the
maximum likelihood estimation \cite{Hradil,StatMeth} to reconstruct the
photon number distribution from the homodyne statistics. This approach
incorporates in a natural way the finite character of the experimental
data sample, as well as physical constraints on the reconstructed
quantities.  Furthermore, compensation for the detector imperfection is
consistently built into the reconstruction scheme, without generating
unphysical artifacts.  Compared to the recent application of the
least--square inversion method to quantum state reconstruction
\cite{OpatWelsPRA97,TanJMO97}, our algorithm is based directly on the
exact statistical analysis of a homodyne experiment.  This
automatically assigns proper statistical weights to experimental
frequencies of homodyne events, and no simplifying assumptions about
the noise present in the finite sample of data are necessary.

We will start with a statistical description of data collected in a
homodyne setup. The phase--averaged probability distribution $p(x)$ of
recording a specific outcome $x$ is a linear combination of the photon
number distribution $\rho_{n}$:
\begin{equation}
p(x) = \frac{1}{2\pi} \int_{0}^{2\pi}
\text{d}\theta \, p(x;\theta) =
\sum_{n=0}^{\infty} A_n(x) \rho_n,
\end{equation}
with coefficients given by the formula \cite{TanJMO97}:
\begin{equation}
\label{Eq:Anx}
A_n(x)= \sum_{m=0}^{n} 
{ n \choose m }
\frac{(1-\eta)^{n-m}\eta^{m}}{\sqrt{\pi}2^m m!}
H_m^2(x) \exp(-x^2),
\end{equation}
where $\eta$ is the detection efficiency and $H_n(x)$ denote
Hermite polynomials. The continuous variable $x$ is divided into bins
of the width $\Delta x$, which we will label with $\nu$. When bins are
sufficiently small, we may approximate the probability $p_\nu$ of
registering the outcome in a $\nu$th bin by:
\begin{equation}
p_\nu = p(x_\nu) \Delta x,
\end{equation}
where $x_\nu$ is the central point of the $\nu$th bin. 

Repeating the measurement $N$ times yields a frequency histogram $
\{k_\nu\}$ specifying in how many runs the outcome has been found in
a $\nu$th bin. The probability of obtaining a specific histogram
$\{k_\nu\}$ is given by the multinomial distribution:
\begin{equation}
\label{Eq:Pknurhon}
{\cal P} ( \{ k_\nu \} | \{ \rho_n \} ) = N! \prod_{\nu} 
\frac{{p_\nu}^{k_\nu}}{k_\nu!},
\end{equation}
where we have explicitly written its dependence on the photon number
distribution $\{\rho_{n}\}$ entering the right hand side via
probabilities $p_\nu$.

When processing the experimental data, we face an inverse problem: given a
certain histogram $\{k_\nu\}$ we want to reconstruct the photon number
distribution $\{\rho_n\}$. The answer given to this problem by the
maximum likelihood method is that the best estimate for $\{\rho_n\}$
maximizes the function defined in Eq.~(\ref{Eq:Pknurhon}), with
$k_\nu$'s treated as fixed parameters obtained from an experiment.
The search for the maximum of the function ${\cal P}$,
called the likelihood function, is {\em a priori} restricted to the
manifold of $\{\rho_n\}$ that describe a possible physical situation.
This guarantees that the reconstructed probability
distribution will be free from unphysical artifacts.

This maximization problem has to be solved by numerical means. For this
purpose we will introduce a cut--off parameter $K$ for the photon
number. The positivity constraints for $\rho_n$ can be satisfied by a
substitution of variables: $\rho_n=y_n^2$. Instead of computing the
likelihood function, it is more convenient to consider its logarithm:
\begin{equation}
{\cal L}(y_0,y_1,\ldots,y_K) = \sum_{\nu} k_\nu \log
\left( \sum_{n=0}^{K} A_{\nu n} y_{n}^{2} \right),
\end{equation}
where we have omitted terms independent of $y_n$, and denoted $A_{\nu n}
= A_{n}(x_\nu) \Delta x$.
The condition that the sum of probabilities $\rho_n$
is equal to one can be taken into account using a Lagrange multiplier
${\cal N}$. Thus we obtain a set of $K+1$ equations:
\begin{eqnarray}
0 & = & \frac{\partial}{\partial y_m} \left(
{\cal L} - {\cal N} \sum_{n=0}^{K} y_{n}^{2} \right)
=
2 y_m \left( \sum_{\nu} \frac{k_\nu}{p_\nu} A_{\nu m} -
{\cal N} \right),\nonumber \\
& & \hspace*{2.15in} m=0,1,\ldots,K. \nonumber \\
& &
\end{eqnarray}
Multiplying these equations by $y_m$ and adding them together yields:
\begin{equation}
{\cal N} = \sum_{\nu} 
\frac{k_\nu}{p_{\nu}}
\sum_{m=0}^{K}
A_{\nu m} y_{m}^{2} = 
\sum_{\nu} k_{\nu} = N,
\end{equation}
that is, the Lagrange multiplier is equal to the total number of
experimental runs.

\begin{figure}
\epsfxsize=3.375in\epsffile{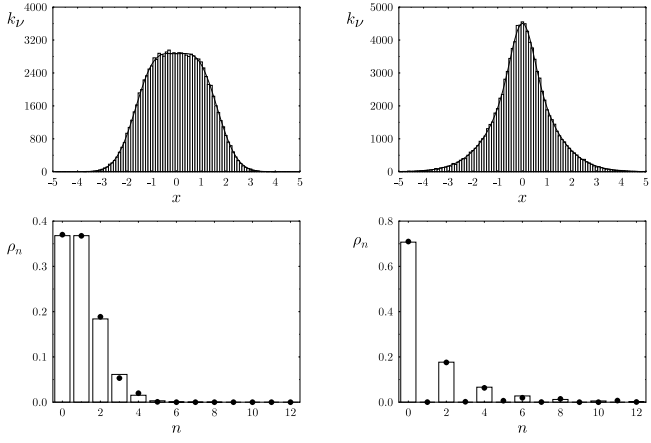}
 
\vspace{0.5cm}
 
\caption{Monte Carlo simulations for 
a coherent state (left) and a squeezed vacuum state (right).
The top graphs depict phase--averaged homodyne statistics
obtained from $N=10^{5}$ runs, superposed on analytical distributions
shown with solid lines. The interval $-5 \le x \le 5$ is divided into
100 bins. The bottom graphs compare the reconstructed photon number
distribution (circles) with analytical values (bars). The cut--off
parameter is $K=19$.}
\label{Fig:Recons}
\end{figure}

The maximization problem reformulated in this way can be treated with
standard numerical procedures. We have performed Monte Carlo
simulations of the homodyne experiment and applied the downhill simplex
method \cite{NumericalRecipes} to reconstruct the photon number
distribution via maximization of the likelihood function. In
Fig.~\ref{Fig:Recons} we depict estimation of the photon number
distribution for a coherent state and a squeezed vacuum state,
both states with the average photon number equal to one. We have
assumed imperfect detection characterized by the efficiency
$\eta=85\%$, which has been taken into account in the reconstruction
process by setting appropriately coefficients $A_n(x_\nu)$ defined in
Eq.~(\ref{Eq:Anx}). The simulated homodyne data are depicted in the top
graphs, and the result of the reconstruction is compared with
theoretical photon number distributions in the bottom graphs. The
even--odd oscillations in the photon number distribution of the squeezed
vacuum state are fully recovered despite losses in the detection process.
Let us note that the maximum likelihood estimation algorithm can be
applied to incomplete data consisting only of selected histogram bins,
provided that they contain enough information to resolve contributions
from different Fock states. This feature may be useful when, for
example, statistics in some bins is corrupted by external noise.

In conclusion, we have presented a method for reconstructing the photon
number distribution from homodyne data via maximization of the
likelihood function.  It allows one to reduce the statistical error by
including in the reconstruction scheme {\em a priori} constraints on the
quantities to be determined.  This method has solid methodological
background and has been derived from the exact statistical description
of a homodyne experiment.

The author is indebted to Professor K.~W\'{o}dkiewicz for numerous
discussions and valuable comments on the manuscript. This research was
supported by the Polish KBN grant No.\ 2P03B~002~14 and by the Foundation
for Polish Science.

\end{document}